\documentclass[11pt]{article}
\usepackage{graphicx}
\begin{document}
\pagenumbering{arabic}

\begin{center}
\title{AMPLITUDE-PHASE ANALYSIS OF COSMIC MICROWAVE BACKGROUND MAPS}
\author{\hspace{0.5cm} P. Naselsky$^{2}$, D.Novikov$^{1,3}$ and 
Joseph Silk$^{1}$. \\
$^1$ Astronomy Department, University of Oxford, NAPL, Keble Road, \\
Oxford OX1 3RH, UK \\
$^2$ Theoretical Astrophysics Center, Juliane Maries Vej 30,
DK-2100 \\
Copenhagen, Denmark \\
$^3$ Astro-Space Center of P.N.Lebedev Physical Institute, 
Profsouznaya 84/32, \\
Moscow, Russia, \\
$^4$ Rostov State University, Zorge 5, Rostov-on-Don, Russia}
\date{}
\maketitle
\end{center}
\begin{abstract}
                                                            
We propose  a novel method for the extraction of unresolved point
sources from
CMB maps. This method is based on the analysis of the phase distribution
of the Fourier components for the observed signal and unlike most
other methods of
denoising  does not require  any significant assumptions about the
expected CMB signal.
The aim of our paper is to show how, using our algorithm, the contribution from
point sources can be separated from the resulting signal on all scales. We
believe that this technique is potentially a very powerful
tool for  extracting  this type of noise from future high resolution maps.

\vspace{0.3cm}

{\it Subject headings:} cosmic microwave background, 
cosmology, statistics, observations.
\end{abstract}

\section{Introduction}
Observations of the Cosmic Microwave Background (CMB) is fundamental
for our understanding  the primordial inhomogeneity of the Universe.
After the successful COBE experiment, attention has been focused
on the investigation of small scale perturbations, that can provide
unique information about the most important cosmological parameters.
One of the major problems in the modern CMB cosmology is to
separate  noise of  various origins (such as dust emission,
synchrotron radiation and
unresolved point sources (see e.g. Banday et al. 1996)) from
the original cosmological signal. Many authors have already applied
various
methods such as  Wiener filtering (Tegmark and Efstathiou 1996, Bouchet
and Gispert 1999),
maximum entropy technique (Hobson et al. 1999),  radical compression
(Bond et al. 1998),
power filtering (Gorski et. al. 1997, Naselsky et. al. 1999) and
wavelet techniques ( e.g. Sanz et al. 1999) to extract noise from the CMB
data.

All of these techniques
have been tested for removing the noise from the real observational data.
It is necessary to note that, for different strategies and for different
experiments, different schemes could be chosen as most appropriate. The
choice
of the algorithm  also depends on the particular type of foreground
emission
to be extracted.

The aim of our paper is to overcome the problem of detecting and
extracting
the
background of unresolved point sources from the original map.
The measured signal in the real observational data is
always smoothed with some
filtering angle $\theta_f$ because of the final antenna beam resolution.
Therefore, unresolved point sources could make a significant contribution
to
the resulting signal on all scales.
This type of noise
should be removed from the original map before any subsequent
analysis is made.

Recently (Cayon et al. 1999) have proposed the use of isotropic wavelets
for removing noise in the form of point sources. Their technique is
based on the fact, that the field in the vicinity of the source
should be in the form of the antenna profile. Unfortunately the Gaussian
CMB field can also form real peaks with the same profile, so that
a lot of 'artificial sources' could be found using this technique.
Besides,
the antenna profile is not necessarily isotropic (indeed, as a rule it
is very anisotropic). Therefore, isotropic wavelets should not be
considered as an absolute cure against such a type of noise.

In this paper we consider an approach, which is based on the distribution
of phases. The idea of using phases of random fields was introduced by
A.Melott et al (1991); Coles and Chiang (2000a,b) for the Large Scale
Structure formation in the Universe. Below we develop the phase-amplitude
analysis method for investigation of the CMB anisotropy and foreground.
The outline of the paper is as follows. In section 2
we briefly review the basic definitions, consider a simulated
one-dimensional
scan of the CMB first with a single point source, then with a background
of such a
sources. In section 3 we generalize our results into two-dimensional maps.
Finally, we suggest an  algorithm for denoising.
In section 4 we
discuss the results and potential of the method for analyzing high
resolution maps.

\section{Point sources in one-dimensional scans.}

In this section we consider 1D CMB scans  with a
background of point sources. This approach could be very
useful for data analysis of one-dimensional experiments
with high resolution (such as  RATAN 600). We extend this discussion
to two-dimensional experiments (such as the new generation of
interferometer experiments) in section 3.
The
investigation of point sources is especially easy in one
dimension, can be easily generalized into two-dimensional maps
and will  help us to understand the advantage of the proposed technique.

\begin{center}
{\bf Definitions}
\end{center}

In 1D the deviation of the temperature from its mean value
$\Delta T=T- \langle T \rangle$ in a scan is
described by the simple Fourier series:

\begin{equation}
\Delta T(\theta)=\sum_k a_k \cos(k\theta)+b_k \sin(k\theta)
\end{equation}
where k is an integer number and $\theta$ can be expressed in terms of
of the real angle on the sky ($\theta_{sky}$) as follows:
$\theta=\frac{\theta_{tot}}{2\pi}\theta_{sky}$. Here $\theta_{tot}$ means
the total length of the scan.

The detected temperature fluctuations $\Delta T$ can as usual be naturally
divided
into two parts: cosmological signal and noise:

\begin{equation}
\Delta T(\theta)=\Delta T^s(\theta) + \Delta T^n(\theta)
\end{equation}
where $s$ and $n$ denote signal and noise respectively.
Therefore, the Fourier transform components $a_k, b_k $ can be also
expressed
as a sum of Fourier decomposition of these two terms:

\begin{equation}
\begin{array}{l}
a_k=a_k^s+a_k^n,\\
b_k=b_k^s+b_k^n.\\
\end{array}
\end{equation}

The statistically isotropic distribution of the
CMB temperature anisotropy is supposed to be in the form of a
random Gaussian field with the power spectrum $P_{CMB}(k)$, which
determined by the appropriate cosmological model. The coefficients
$a_k^s, b_k^s$  depend on the spectrum of the CMB, the antenna filtering
function
$\widetilde{F}(\theta-\theta^*,\theta_f)$ and the actual realization
of the random Gaussian process on the sky. In general, they obey the
formulae:
$ \langle a_k^sa_{k'}^s \rangle =
\langle b_k^sb_{k'}^s \rangle =\delta_{kk'}F(k,k_f)P_{CMB}(k)$.
Here, $F(k,k_f)$ is the
Fourier transform of the filtering function and $\theta_f$ is the antenna
resolution angle. $k_f$ is a wavenumber which corresponds to this
resolution:
$k_f=1/\theta_f$. In our simulations we use the usual expression for
$a_k,b_k$:

\begin{equation}
\begin{array}{l}
a_k^s=\alpha_kF^{\frac{1}{2}}(k,k_f)P_{CMB}^{\frac{1}{2}}(k),\\
b_k^s=\beta_kF^{\frac{1}{2}}(k,k_f)P_{CMB}^{\frac{1}{2}}(k),\\
\end{array}
\end{equation}
where $\alpha_k,\beta_k$ are independent Gaussian numbers with zero mean
and
unit dispersion.

In this paper we consider the noise in the form of isolated unresolved
point sources.
This  means that the average distance between sources is larger than
the resolution scale $\theta_f$.
Therefore, the shape of the 'noise' field around the point source
determined by the
filtering function F:

\[
\Delta
T_n(\theta)=\int\sum\limits_{j=1}^{N_{ps}}\gamma_j\delta(\theta^*-\theta)
\widetilde{F}^{\frac{1}{2}}(\theta^*-\theta,\theta_f)d\theta^*=
\]
\begin{equation}
=\sum\limits_{j=1}^{N_{ps}}\gamma_j\widetilde{F}^{\frac{1}{2}}
(\theta-\theta_j,\theta_f)
\end{equation}
where $\gamma_j,\theta_j$ are the amplitude and the position of the j-th
point source, respectively, and
$N_{ps}$ is the total number of point sources in the considered scan.
According to equation [5], the Fourier components of the noise can be
described by the following very
simple and convenient formulae:

\begin{equation}
\begin{array} {l}
a_k^n=\sum\limits_{j=1}^{N_{ps}}\gamma_j\cos(k\theta_j)F^{\frac{1}{2}}(k,k_f),\\
b_k^n=\sum\limits_{j=1}^{N_{ps}}\gamma_j\sin(k\theta_j)F^{\frac{1}{2}}(k,k_f).\\
\end{array}
\end{equation}

For further investigation we have to introduce the phase: $\varphi_k$ of
the k-th harmonic.
Using equations [1,3] one can write:

\begin{equation}
\varphi_k=\arctan\left[\frac{b_k}{a_k}\right]=
\arctan\left[\frac{b_k^s+b_k^n}{a_k^s+a_k^n}\right]
\end{equation}

If the resulting field at the scales $k$ is dominated by the Gaussian CMB
signal ($S_k/N_k>>1$),
then $\varphi_k \approx arctg(b_k^s/a_k^s)$. In this case the phases of
the  k-th
harmonics are random independent uncorrelated  values, uniformly
distributed from 0 to $2\pi$.
On the other hand, if the
signal at these scales is much smaller than the noise, then the
distribution of phases is
determined by the positions and amplitudes of point sources on the scan.
In Fig.1,
we present the spectrum of CMB in one dimension $P_{CMB}(k)$ for the
standard CDM model
together with the spectrum of point sources. Both spectra are smoothed
with the Gaussian
filtering function $F(k,k_f)=\exp(-\frac{k^2}{2k_f^2})$. It is well known
that the CMB
signal disappears when $k$ becomes larger than some value $k_d$. This value
corresponds to
the damping scale of the CMB fluctuations. Therefore, at the small scales
the resulting field
is dominated by the noise.
Note that $k_d$ should not be necessary interpreted as the damping scale.
Roughly speaking, this is the scale where noise from sources becomes
larger
then the CMB signal.

\parbox{4.5in}{
      {\includegraphics[scale=0.55,width=4.5in,totalheight=2.5in]{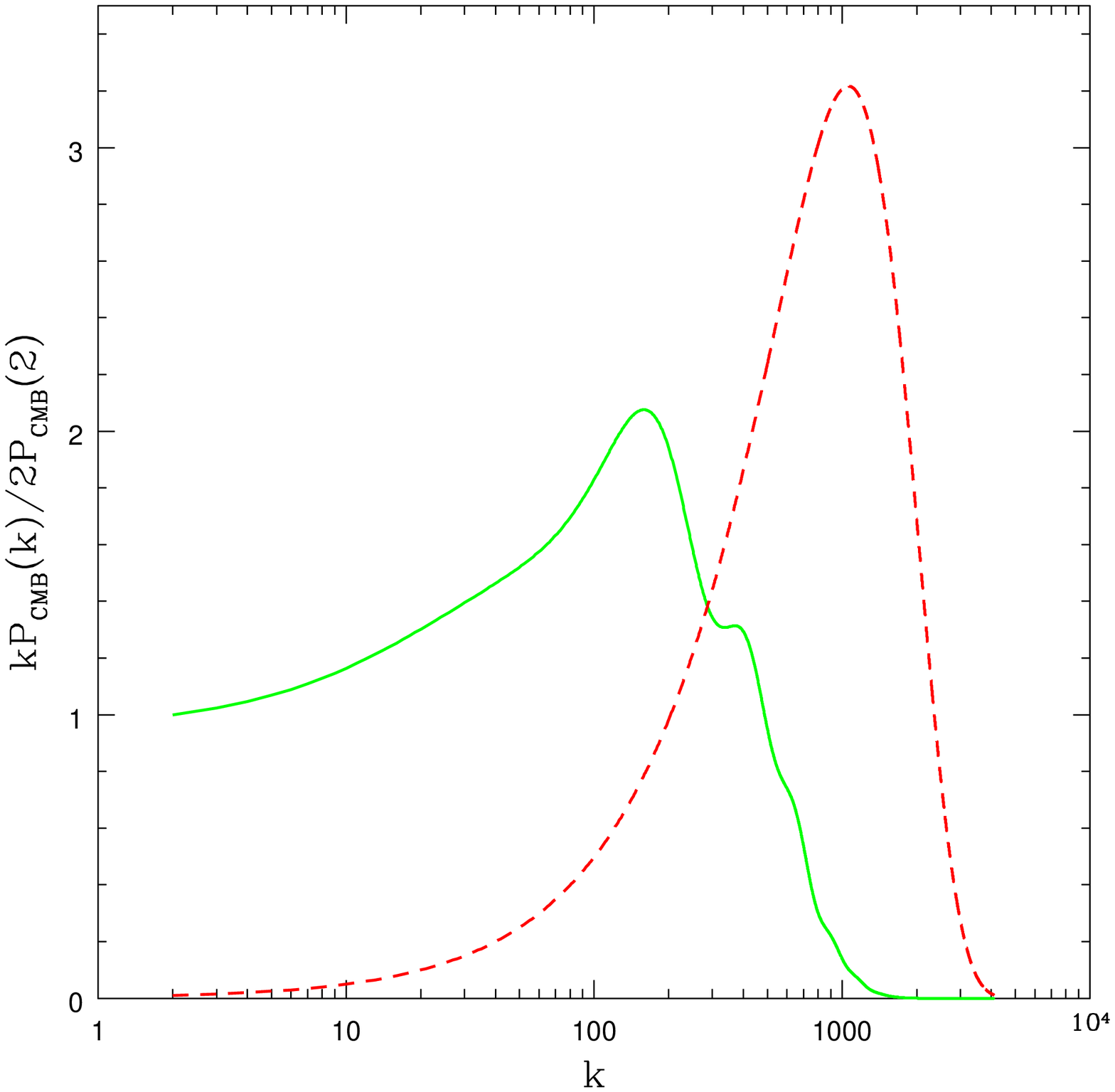}
}

\
      {\small {\bf{Fig.~1} \ }
      { The power spectrum of the CMB (solid line) for one-dimensional 
     ($360^o$) 
     scan together with the spectrum of point sources (dashed line).

}}}

\vspace{1cm}

It is easy to see, from equations [4,6,7], that the process of smoothing
does not change the
phases of the primordial signal. The filtering function $F(k,k_f)$ has
simply disappeared
from the right hand side of the equation [7]. Therefore, if $k_f>k_d$,
 we have the
possibility of measuring the phases  only for high $k$ values of the noise.
Below we describe how
the information about the phase distribution for high values of $k$ can be
used for very
precise detection and extraction of the contribution from the sources for
all values of
$k$

\begin{center}
{\bf Detection of a single point source}
\end{center}

Let us consider the simplest example by  dealing  with a single
unresolved
point source on the scan. In order to remove the contribution from
this source, we have to know its precise location $\theta_1$ and
amplitude $\gamma_1$ (see Fig.2).

\parbox{4.5in}{
      {\includegraphics[scale=0.55,width=4.5in,totalheight=4.5in]{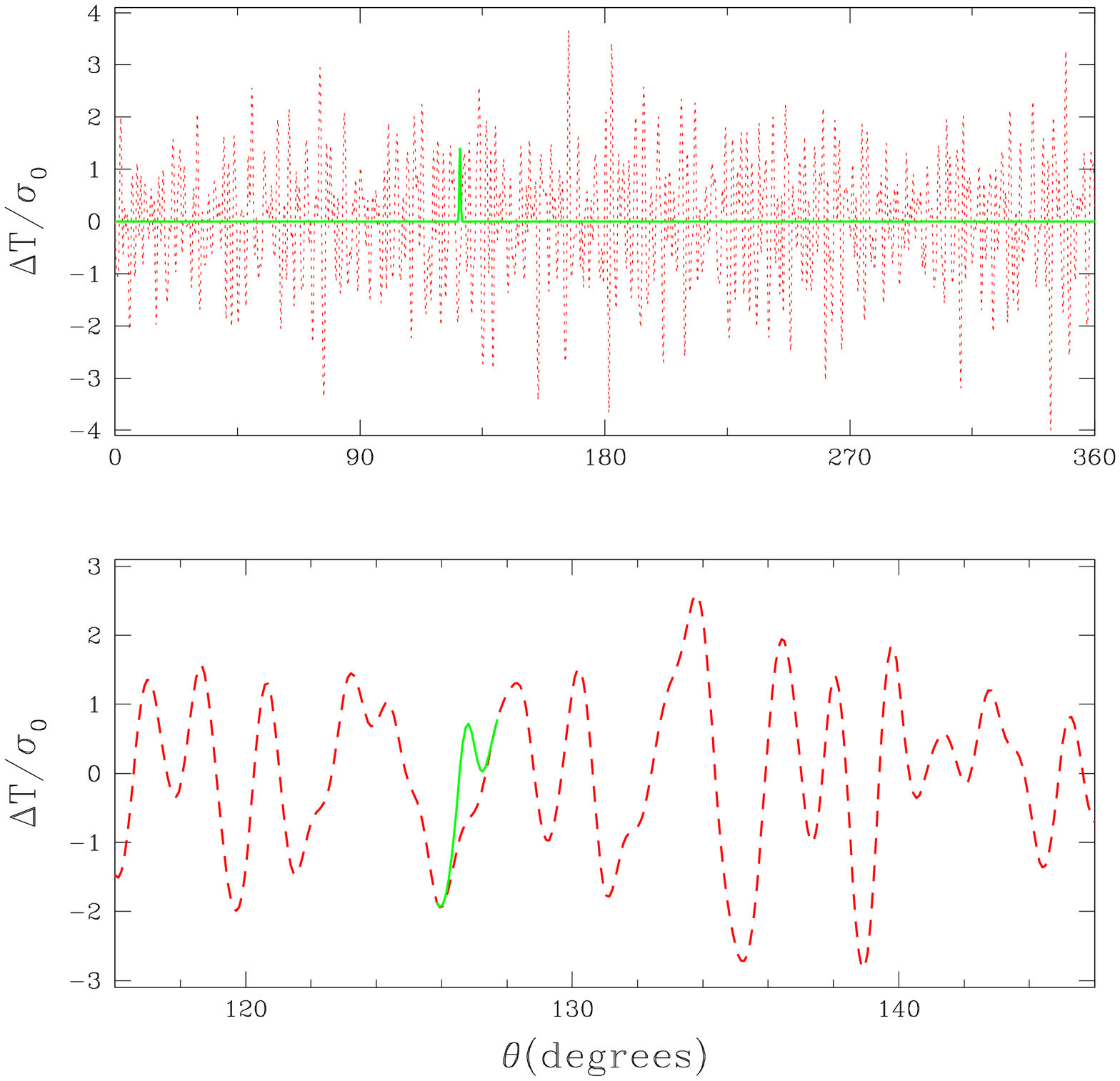}
}

\
      {\small {\bf{Fig.~2} \ }
      { Upper panel: simulated CMB field on $100^o$ scan ($1^o$
corresponds 
to $\approx 0.03^o$ on the sky) (dashed line) and the field from a 
single point source (solid line). Lower panel: the same as the upper one,
but
with better resolution. The field in the vicinity of the point source
behaves
like an ordinary Gaussian fluctuation. 
}}}

\vspace{1cm}

The contribution from this source to the resulting field according to
equation [5] is then:

\begin{equation}
\Delta T_n=\sum\limits_{k=1}^{k_{max}}\gamma_1\cos(k(\theta-\theta_1))
F^{\frac{1}{2}}(k,k_f)
\end{equation}
where $k_{max}$ is the maximum value of $k$ that can be detected in the
experiment.

As has been already mentioned, for $k$ larger then some value $k_d$,
phases $\varphi_k$ are just the phases of the point source. From equations
[6,7], we obtain:

\begin{equation}
\varphi_k=mod_{2\pi}(k\theta_1)
\end{equation}

It suffices to have only two phases (for example $\varphi_k$ and
$\varphi_{k+1}$,
$k>k_d$) to find the location of the source $\theta_1$:

\begin{equation}
\theta_1=\varphi_{k+1}-\varphi_k
\end{equation}
In Fig.3 we show the behavior of the phases $\varphi_k$, $1<k<k_{max}$
together with the phases of the source. For small values of $k$: $k<<k_d$
the phases are distributed uniformly and at large $k$ we can definitely
see
the regular structure that is consistent with equation [9].

\parbox{4.5in}{

{\includegraphics[scale=0.55,width=4.5in,totalheight=4.5in]{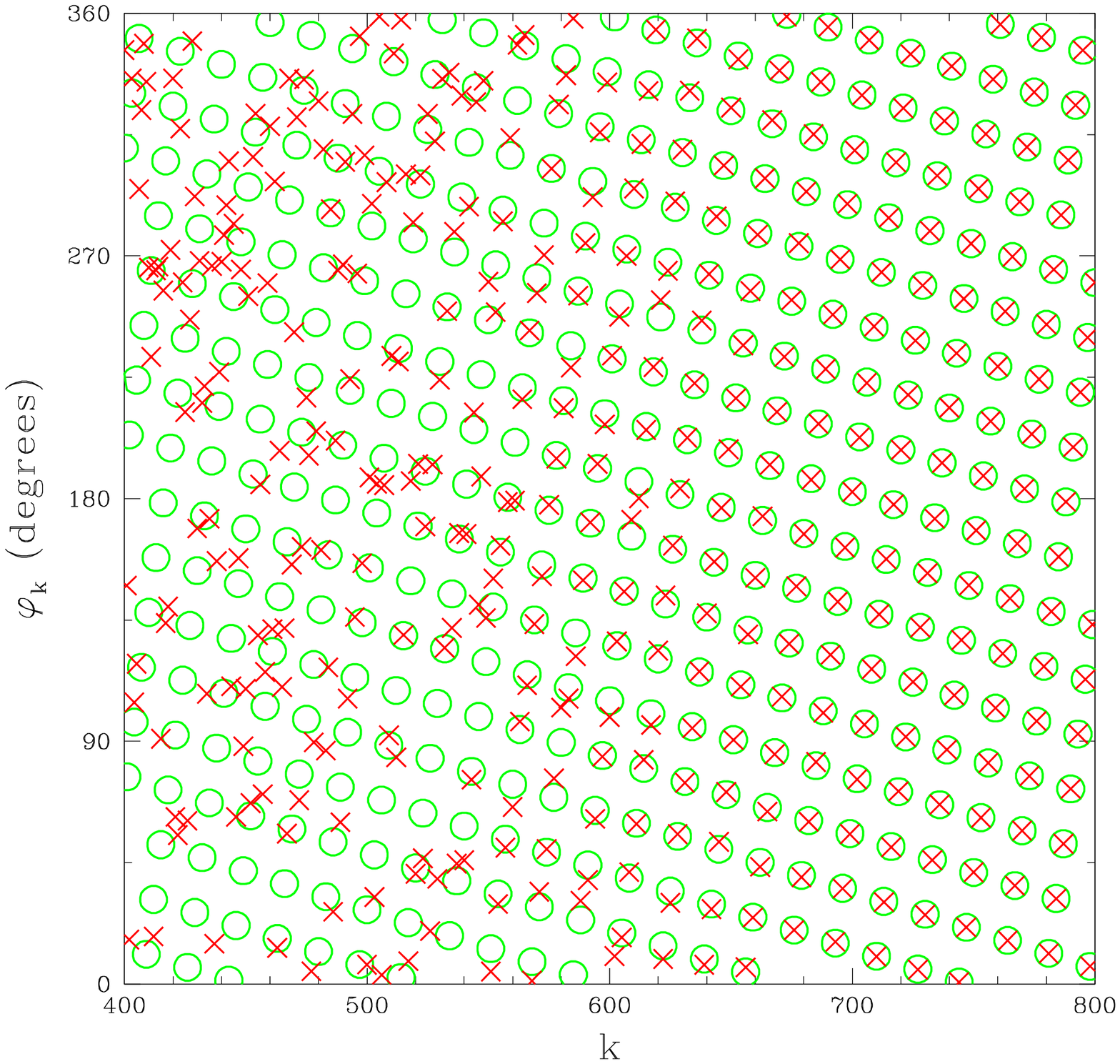} }

\
      {\small {\bf{Fig.~3} \ }
      { The phases of the point source (circles) and phases of the
resulting signal: CMB + Point source (crosses).

}}}

\vspace{1cm}

In Fig.4 we also show the positions of maxima for all harmonics. Location
of the maxima for the k-th harmonic can be found by the formulae:

\begin{equation}
\theta_{max}^k=\frac{\varphi_k+2\pi*n}{k}
\end{equation}
where n is an integer number. The straight vertical line points to
the location of the source because one of the maxima in each harmonic
is coincident with this location.

\parbox{4.5in}{

{\includegraphics[scale=0.55,width=4.5in,totalheight=3.5in]{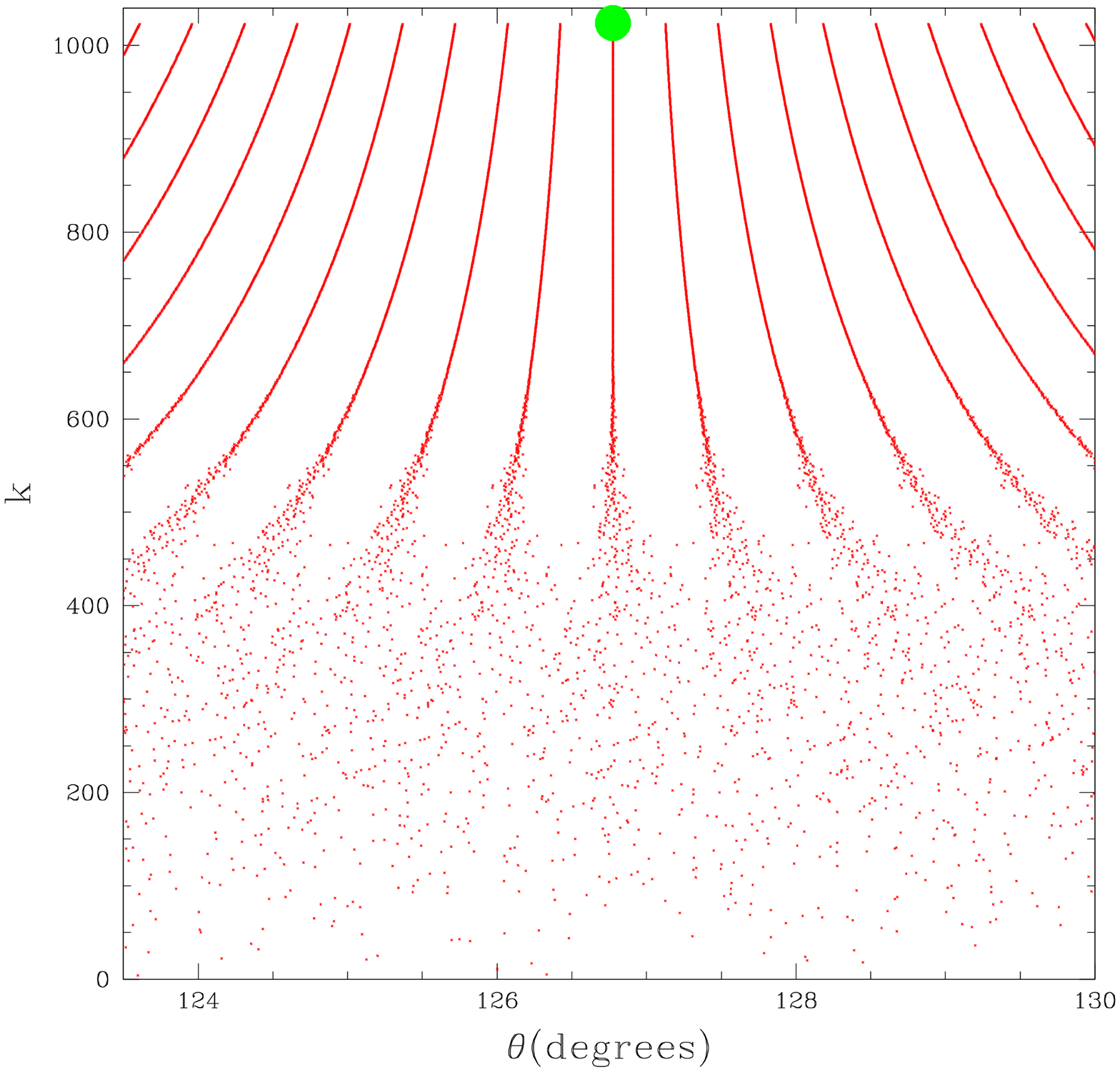} }

\
      {\small {\bf{Fig.~4} \ }
      { Positions of maxima for each harmonic. Each point
represents the positions of maxima for the k-th harmonic. For small k
they are distributed uniformly (according to the Gaussian distribution
of the CMB). The large dot shows the location of the source.  

}}}

\vspace{1cm}

The remaining part of the problem is to find the amplitude - $\gamma_1$.
Let us defined the field $\Delta T^{k_d}(\theta)$ as a part of the field
$\Delta T(\theta)$ that consists  only of the high harmonics:

\begin{equation}
\Delta T^{k_d}(\theta)=\sum\limits_{k=k_d}^{k_{max}}a_k \cos(k\theta)+b_k
\sin(k\theta)
\end{equation}
Using the formulae [8], we now can write down the obvious relation:
\begin{equation}
\gamma_1= \Delta T^{k_d}(\theta_1)/\sum\limits_{k=k_d}^{k_{max}}F(k,k_f)
\end{equation}

Therefore, according to [3,6], we have found the contribution from this
source to all harmonics from $k=1$ to $k=k_{max}$.

\begin{center}
{\bf Background of point sources}
\end{center}

In this subsection we generalize our algorithm to the case where  there are
an unknown number of point sources in the considered scan. In a situation
like this, we have to find not only positions and amplitudes of each
source but also the total number of them: $N_{ps}$.

We believe that many different techniques based on the results
of the previous subsection could be proposed to solve this
problem. We suggest a simple iteration scheme.
As has been already noticed above, we can consider the
field $\Delta T^{k_d}$, which consists  only of high harmonics.
Therefore, only point sources make a contribution to this
field:
\begin{equation}
\begin{array} {l}
\Delta T^{k_d}(\theta)=\sum\limits_{k=k_d}^{k_{max}}a_k^n \cos(k\theta)
+b_k^n \sin(k\theta)=\\
=\sum\limits_{j=1}^{N_{ps}}
\gamma_j\sum\limits_{k=k_d}^{k_{max}}
F^{\frac{1}{2}}(k,k_f)\cos(k(\theta-\theta_j))
\end{array}
\end{equation}
We now introduce the filter function $L(k)=F(k-k_d,k_l)/F(k,k_f)$
and consider the filtered field:
\begin{equation}
\Delta \widetilde{T}_k^{k_d}=\Delta {T}_k^{k_d}L^{\frac{1}{2}}(k)
\end{equation}
According to [14,15], one can write:

\begin{equation}
\begin{array} {l}
\Delta \widetilde{T}^{k_d}(\theta)=\\
\sum\limits_{j=1}^{N_{ps}}\gamma_j

\cos(k_d(\theta-\theta_j))\sum\limits_{k=1}^{kmax-k_d}
\cos(k(\theta-\theta_j))F(k,k_l)^{\frac{1}{2}} \\

%\hspace{1cm}
-\sin(k_d(\theta-\theta_j))\sum\limits_{k=1}^{kmax-k_d}
\sin(k(\theta-\theta_j))F(k,k_l)^{\frac{1}{2}}

\end{array}
\end{equation}
If we can put $k_l<<k_d<<k_{max}$, then the second term on the right hand
side of equation [16] is small and:

\begin{equation}
\Delta \widetilde{T}^{k_d}(\theta) \approx
\sum\limits_{j=1}^{N_{ps}}\gamma_j
\sum\limits_{k=1}^{kmax-k_d}
\cos(k(\theta-\theta_j))F(k,k_l)^{\frac{1}{2}}
\end{equation}
This equation is very close to [8] and, therefore, the procedure of
filtering gives us the possibility of 'localizing' the field in the
vicinity
of a point source.

In reality equation [17] is not quite correct because $k_d$ and $k_{max}$
are values of approximately the same order and, therefore, peaks,
that are more or less close to each other can interfere (fig. 4).
This is the reason
why we choose the iteration technique to remove point sources.

We propose the following algorithm. Let us construct the field
$\Delta \widetilde{T}_o^{k_d}=\Delta \widetilde{T}^{k_d}$
and find its highest maximum. This
maximum most probably corresponds to the most powerful isolated
point source on the scan. The position and value of this maximum
give us the location $\theta_1$ and the amplitude $\gamma_1$ (eq[13])
of this source.
After that, we construct
the field $\Delta \widetilde{T}_1^{k_d}$ 'without' this point source:

\begin{equation}
\Delta \widetilde{T}_1^{k_d}(\theta)=\Delta \widetilde{T}_0^{k_d}(\theta)-
\gamma_1\sum\limits_{k=k_d}^{kmax}
\cos(k(\theta-\theta_1))L(k,k_l)^{\frac{1}{2}}.
\end{equation}
The contribution from this source to the field and its interference
with other sources is now removed. This allows us to find more
precisely the next highest maximum. Therefore, we apply the same procedure
to the field $\Delta \widetilde{T}_1^{k_d}$ and find $\theta_2$,
$\gamma_2$
and so on (Fig.5).

\parbox{4.5in}{
      {\includegraphics[scale=0.55,width=4.5in,totalheight=4.5in]{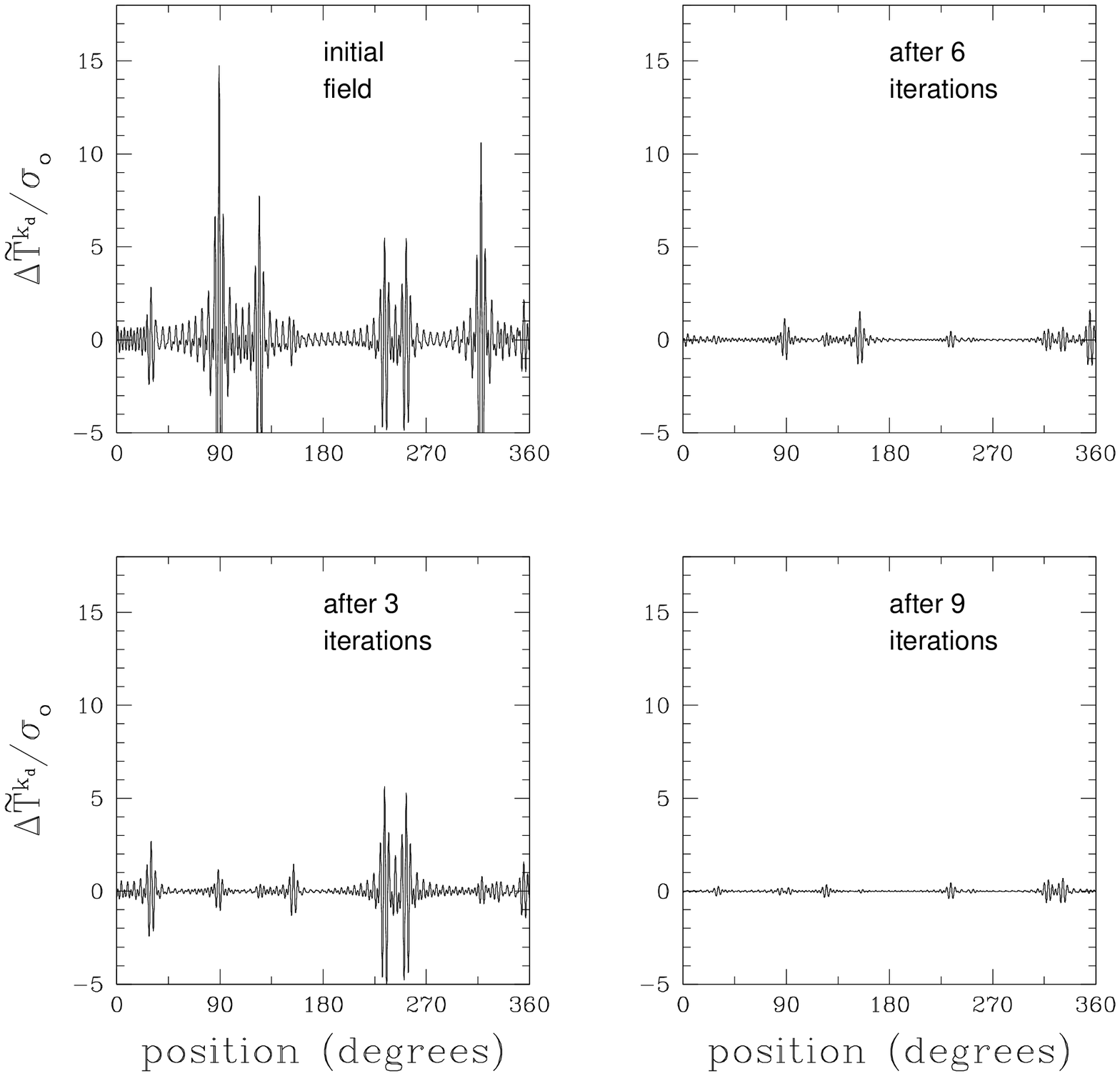}
}

\
      {\small {\bf{Fig.~5} \ }
      { The iteration scheme. Each panel represents the residuals
 $\Delta \widetilde{T}_i^{k_d}$ from
the initial field $\Delta \widetilde{T}_o^{k_d}=
\Delta \widetilde{T}^{k_d}$ after the i-th iteration. 

}}}

\vspace{1cm}

We perform these iterations until the dispersion $\sigma_i^{k_d}$
($(\sigma_i^{k_d})^2=\langle(\Delta \widetilde{T}_i^{k_d})^2 \rangle$)
becomes significantly smaller then $\sigma_o^{k_d}$ (Fig.6).
The total number of iterations that is needed to significantly reduce
the initial dispersion gives us approximately the number of point sources
$N_{ps}$ and each iteration gives the location $\theta_i$ and the
amplitude
$\gamma_i$ of the i-th source.
Note, that
\begin{equation}
(\sigma_i^{k_d})^2=\sum\limits_{j}\gamma_j^2,
\hspace{1cm}\gamma_j<\gamma_i
\end{equation}
and roughly speaking, in Fig.7 we can see the cumulative distribution
of point sources over the power $\gamma$.
Finally, since we have the position $\theta_i$ and   amplitude $\gamma_i$,
the contribution
to the field from all point sources may be
removed in the same manner, as was done for a single point
source in the previous subsection.

\parbox{4.5in}{
      {\includegraphics[scale=0.55,width=4.5in,totalheight=3.5in]{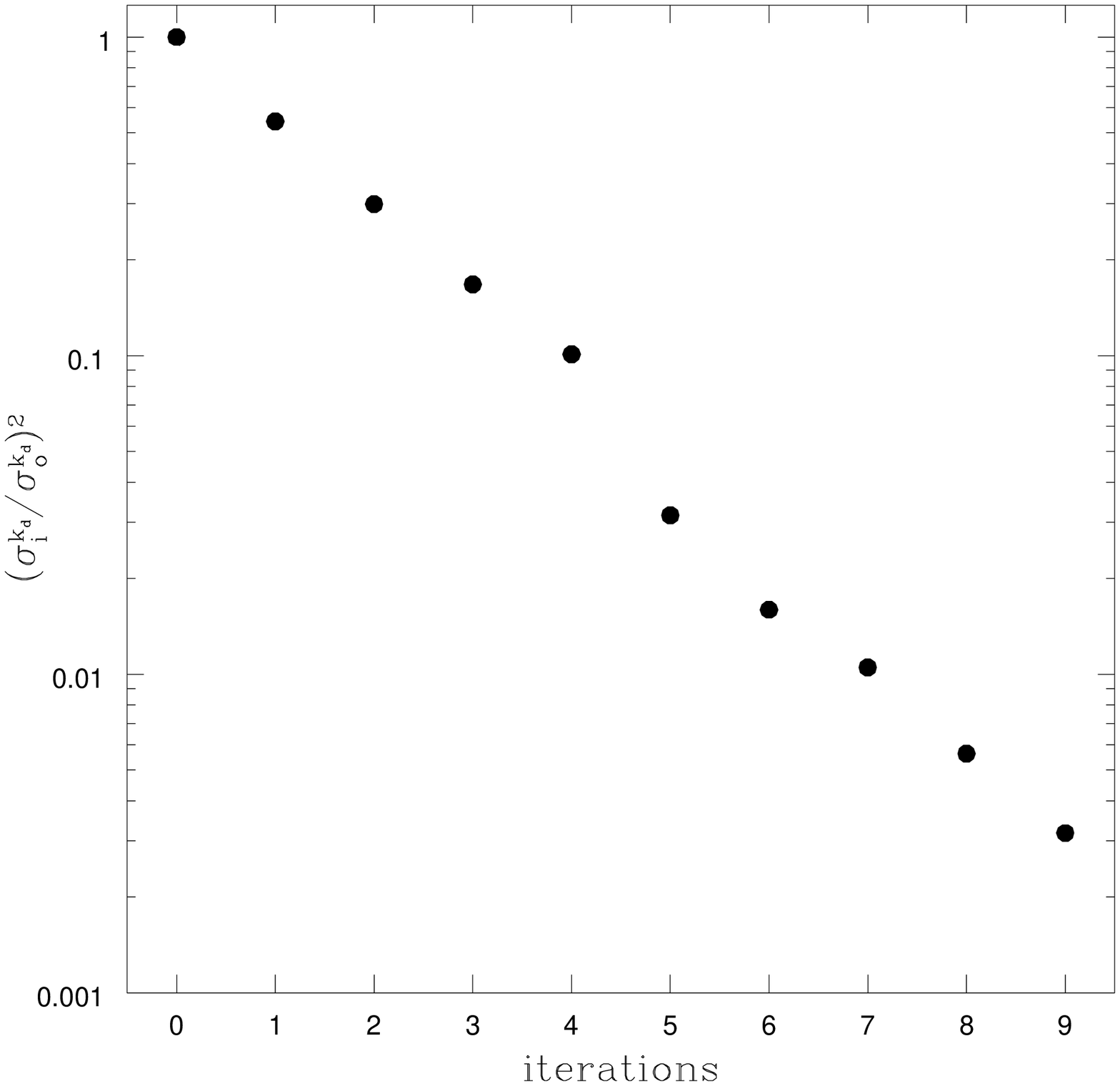}
}

\
      {\small {\bf{Fig.~6} \ }
      {The decrease of the dispersion for 
$\Delta \widetilde{T}^{k_d}$ with each iteration.

}}}

\parbox{4.5in}{
      {\includegraphics[scale=0.55,width=4.5in,totalheight=3.5in]{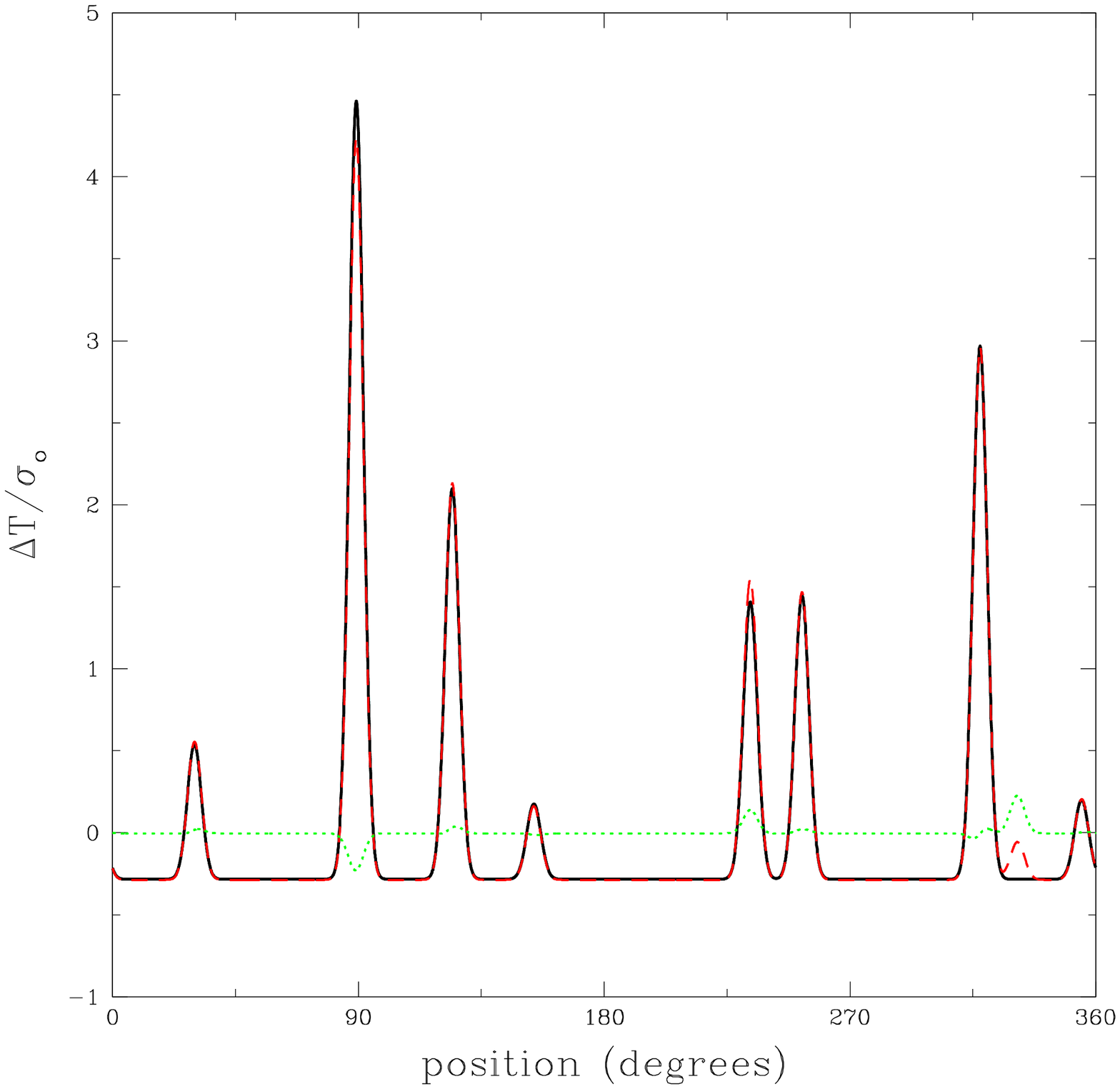}
}

\
      {\small {\bf{Fig.~7} \ }
      { The final result. The initial field of point sources (solid line),
restored field by our method (dashed line), and residuals (dotted line).

}}}

\newpage

\vspace{1cm}

\section{Point sources in two dimensions.}

In this section we briefly describe our results in two dimensions.
Without loss of generality we may consider a small region of the
sky and assume that the geometry is approximately flat. Under
this assumption, the part of the detected signal which is determined
by the noise associated with  $N_{ps}$ point sources can be represented
according to the previous section by writing:

\begin{equation}
\begin{array} {l}
\Delta T_n(\vec{x})=\sum\limits_{\vec{k}}a_k^n \cos(\vec{k}\vec{x})
+b_k^n \sin(\vec{k}\vec{x})=\\
\sum\limits_{j=1}^{N_{ps}}
\gamma_j\sum\limits_{\vec{k}}
F^{\frac{1}{2}}(\vec{k},|\vec{k}_f|)\cos(\vec{k}(\vec{x}-\vec{x}_j))
\end{array}
\end{equation}
where $\vec{x}_j$ is the position of the j-th point source in the
Cartesian
coordinate system and $|\vec{k}_f|$ corresponds to the antenna resolution.
Analogously to the one-dimensional case, this field should be filtered with
some appropriate  function.
The convenient filter function $L(\vec{k})$ that we use in this case is as
follows:

\begin{equation}
\begin{array} {l}
L(\vec{k})=F^{-1}(\vec(k),|\vec{k}_f|) \hspace{0.8cm} if \hspace{0.2cm}
|\vec{k}_d|<|\vec{k}|<|\vec{k}_{max}|,\\
L(\vec{k})=0 \hspace{1.1cm} if \hspace{0.4cm} |\vec{k}|<|\vec{k}_d|
\hspace{0.2cm} or
\hspace{0.2cm} |\vec{k}|>|\vec{k}_{max}|.
\end{array}
\end{equation}
According to [15,20,21] one can write:

\begin{equation}
\Delta \widetilde{T}^{k_d}(\vec{k})= \sum\limits_{j=1}^{N_{ps}}\gamma_j
\sum\limits_{|\vec{k}|=|\vec{k}_d|}^{|\vec{k}|=|\vec{k}_{max}|}
\cos(\vec{k}(\vec{x}-\vec{x}_j))
\end{equation}
Therefore, the filtered function $\Delta \widetilde{T}^{k_d}(\vec{k})$ at
the point $\vec{x}=\vec{x}_j$
has a peak with amplitude equal to the power of j-th point source times the
number of modes that we
can use for data analysis in the appropriate experiment.

In our simulations of the signal+noise we use the standard CDM model and
background of
100 point sources randomly distributed over the $10^o \times 10^o$ map.
Without loss of
generality we use the
simple symmetric Gaussian antenna profile. All these calculations, of
course,  could be done for any arbitrary antenna beam.
In Fig.8  we show the map of the CMB
together with the maps of noise, CMB+noise and the filtered map of
CMB+noise. The last
one shows us more or less clearly the positions and powers of point
sources. The significant
anisotropy that appears in the last map occurs for  the following reason.
According to
formulae [21] we use only the set of harmonics $k_1,k_2$, that obey the
relation
$k_1^2+k_2^2>k_d^2$. Therefore, the number of horizontal and vertical waves is
larger than the number of waves in any other direction. (This problem does not
occur  if we use
spherical harmonics $Y_l^m$ with $l>l_d$).

We apply the same iteration technique as in the previous section and
therefore separate
noise from the cosmological signal (Fig.9). It is necessary to note that
each iteration
removes  an appropriate point source  at the beginning of this
process  for the
most powerful and separated sources.
For the weaker sources, additional iterations are needed.
The signal from the j-th source
decreases as
$\approx \frac{\gamma_j}{((\vec{k}_{max}-\vec{k_d})\delta r)^2}$, where
$\delta r$
is the distance from the peak (in one dimension this dependence is
linear).  

\newpage
\textheight 8.5in
\hoffset-1.5in

\begin{center}	
\parbox{7.5in}{
  \includegraphics[scale=0.8,width=7.5in,totalheight=7.5in]{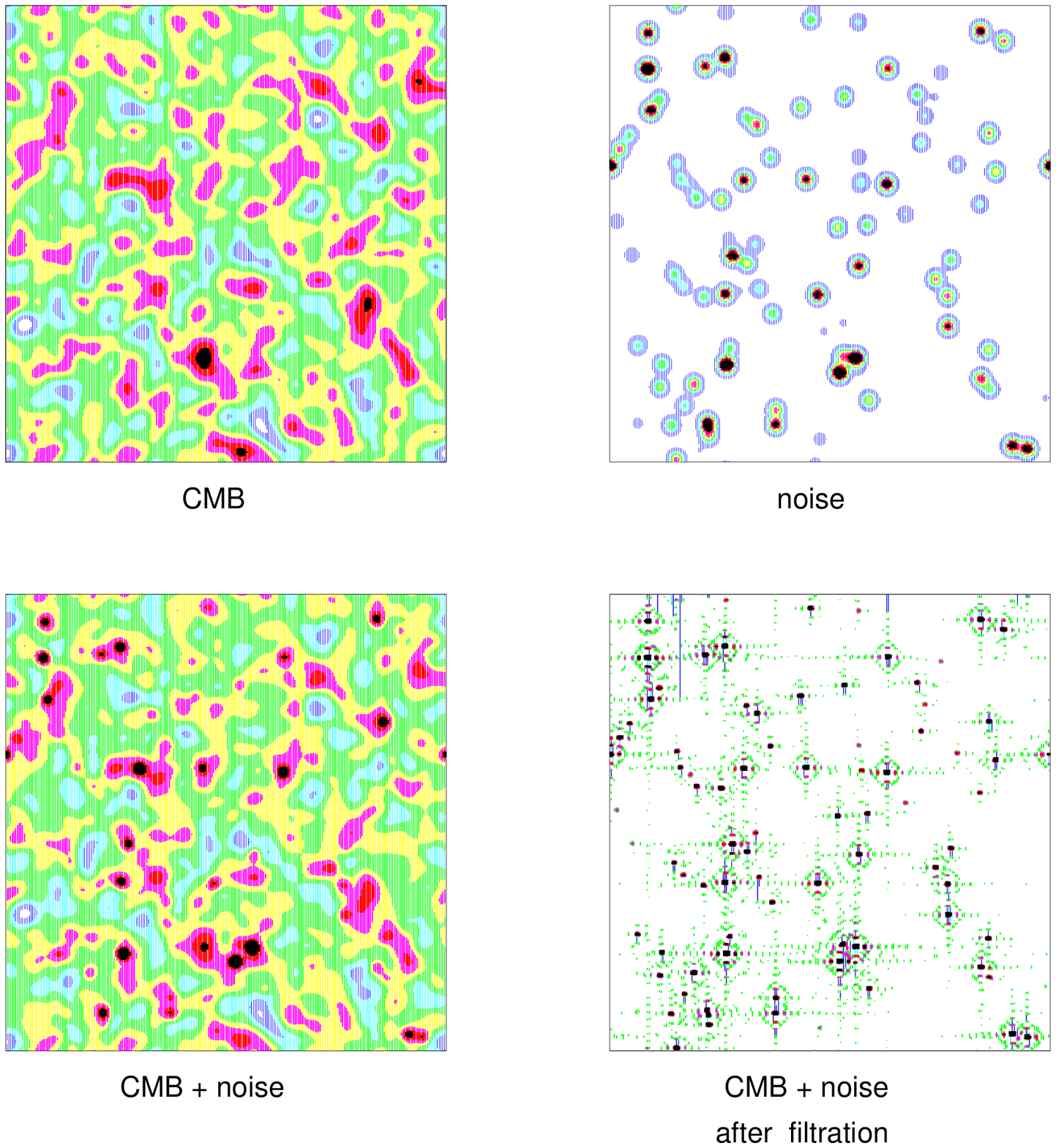}

\begin{center}	
      	{\small {\bf{Fig.~8} \ }
      	{Simulated sky maps of $10^o \times 10^o$.

        }}
\end{center}
}
\end{center}

\begin{center}	
\parbox{7.5in}{
  \includegraphics[scale=0.8,width=7.5in,totalheight=7.5in]{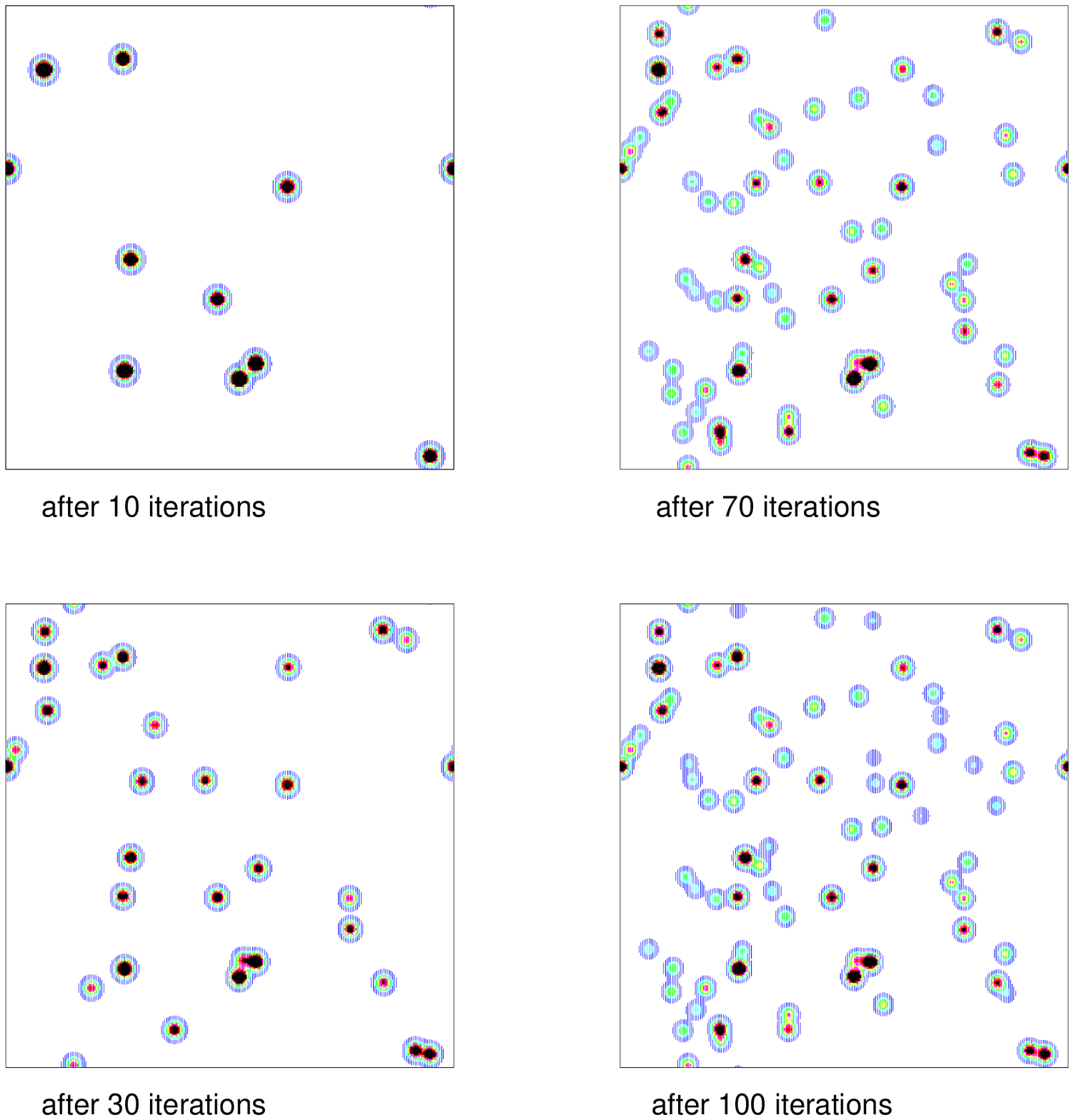}

\begin{center}	
      	{\small {\bf{Fig.~9} \ }
      	{Noise maps (i.e. removed sources) after different numbers of iterations.
The  size and shading of each source  is proportional to its amplitude.

        }}
\end{center}
}
\end{center}
\newpage
\textheight 21cm
\hoffset-0in

This
affects  the neighboring peaks and can change their
amplitudes. Therefore,
this approximation works if $((\vec{k}_{max}-\vec{k_d})\delta r_{ij})^2
<<\frac{\gamma_i}{\gamma_j}$, where $r_{ij}$ is the separation between
the i-th and j-th
peaks (the i-th peak is the closest to the j-th peak). Otherwise, the
amplitudes that have been
found
in each iteration would not correspond to  the powers of the sources and
we therefore
have to perform a
number of iterations that is larger than the number of sources.

\section{Conclusions}

In this paper we present a powerful method for extraction of unresolved
point sources from  future high resolution CMB maps (such as  MAP, Planck,
VSA, CBI, DASY, AMI and RATAN 600).
Our  method is based on the distribution of phases.
The most important advantage of our technique is that we do not  make any
strong
assumptions about the expected CMB signal as well as about the antenna
profile.
Most  other techniques use the  estimated power spectrum of the
CMB and noise
before the
data analysis is implemented (e.g. Wiener filter) or they require special
assumptions about the antenna profile (e.g., wavelets techniques).
It is worth stressing that, for example, assumptions about the CMB power
spectrum
can lead to incorrect interpretations of the observational data. Roughly
speaking, by making such assumptions, one runs the risk of generating the
result one wants and any discrepancies are consider to be  errors.

Our algorithm is numerically very efficient. It is a linear
algorithm and  requires $Nln(N)\times N_{ps}$ operations, where N is the
number of
pixels. Therefore it can be
easily applied to the analysis of large data sets.
  We have demonstrated  the accuracy which can be achieved  using
our algorithm to remove the contribution from  point sources on all
scales.
We believe that this technique is potentially a very powerful
tool for  extracting this type of noise from future high resolution maps.

\begin{center}
{\bf Acknowledgments}
\end{center}
We are very grateful to I.Novikov and A.Doroshkevich for discussions
and P.Coles and R.Scherrer for informative communications.
This investigation was partly supported by INTAS under grant number
97-1192, by RFFI under grant 17625 and by Danmarks Grundforkskningfond
through its support for TAC.

\begin{center}
{\bf References}
\end{center}
.\\
Banday, A.J., Gorski, K.M., Bennett, C.L., Hinshaw, G.,
              Kogut, A., \&
              Smoot, G.F., ApJ. Letters, {\bf 468}, 85, 1996\\
Bond, J.R., A.N.Jaffe and  L.Knox, Phys. Rev. D. 57,
		2117, 1998.\\
Bouchet, F.R. \& Gispert, R. 1999, astro-ph/9903176\\
Cayon, L., Sanz, J.L., Barreiro., R.B., Martinez-Gonzalez, E.,
              Vielva, P., Toffolatti, L., Silk, J., Diego, J.M. and
		F. Argueso astro-ph/9912471\\
Coles, P. and L.Y.Chiang, MNRAS, {\bf 311}, 809, 2000a.\\
Coles, P. and L.Y.Chiang, Nature, {\bf 406}, 376-378, 2000b.\\
Hobson, M.P., Barreiro, R.B., Toffolatti, L., Lasenby, A.N.,
              Sanz, J.L., Jones, A.W. \& Bouchet, F.R. 1999, MNRAS, 306, 232.\\
Gorski, K.M., Proceedings of the 31-st Recontres de Marion
              Astrophysics Meeting, p. 77, 1997, astro-ph/9701191.\\
Guiderdony, B. 1999, astro-ph/9903112\\
Melott, A., S. Shandarin and R. Scherrer, ApJ. {\bf 377},
		79, 1991.\\
Novikov, D.I., Naselsky, P.D., Jorgensen, H.E., Christensen,
 P.R., Novikov, I.D., Norgaarrd-Nielsen, H.U., astro-ph/0001432\\
Sanz, J.L., Barreiro, R.B., Cayon, L., Martinez-Gonzalez, E.,
              Ruiz, G.A., Diaz, F.J., Argueso, F., Silk, J., and
              L. Toffolatti, 1999, astro-ph/9909497\\
Tegmark,M. \& Efstathiou, G. 1996, MNRAS, 281, 1297.

\end{document}